\documentclass[trackchanges]{aastex701}

\def\deg{$^{\circ}$}
\renewcommand\arcmin{$^\prime$}
\renewcommand\arcsec{\mbox{$^{\prime\prime}$}}

\newcommand{\kms}{km~s$^{-1}$}

\newcommand{\Ha}{$\rm H\alpha$}

\newcommand{\hi}{{H{\sc i}}}
\newcommand{\htwo}{H$_2$}

\newcommand{\x}{$\times$}
\newcommand{\about}{$\sim$}
\newcommand{\Msun}{M$_\odot$}
\newcommand{\Msunyr}{M$_\odot$~yr$^{-1}$}

\newcommand{\Mhi}{$M_{\rm HI}$}

\newcommand{\Mhtwo}{$M_{\rm H_2}$}
\newcommand{\Mst}{$M_\star$}

\begin{document}

\title{MAUVE-MUSE: A Star Formation-driven Outflow Caught in the Act of Quenching the Stripped Virgo Galaxy NGC 4064}

\author[orcid=0009-0009-1440-9206,gname= 'Amy', sname='Attwater']{Amy Attwater}
\affiliation{International Centre for Radio Astronomy Research, University of Western Australia, 7 Fairway, Crawley, 6009, Western Australia, Australia}
\email[show]{amy.attwater@research.uwa.edu.au}  

\author[orcid=0000-0002-7625-562X, gname=Barbara,sname=Catinella]{Barbara Catinella}
\affiliation{International Centre for Radio Astronomy Research, University of Western Australia, 7 Fairway, Crawley, 6009, Western Australia, Australia}
\email{barbara.catinella@uwa.edu.au}

\author[orcid=0000-0002-7422-9823, gname=Luca,sname=Cortese]{Luca Cortese}
\affiliation{International Centre for Radio Astronomy Research, University of Western Australia, 7 Fairway, Crawley, 6009, Western Australia, Australia}
\email{luca.cortese@uwa.edu.au}

\author[orcid=0000-0003-4932-9379, gname=Timothy,sname=Davis]{Timothy Davis}
\affiliation{Cardiff Hub for Astrophysics Research \& Technology, School of Physics and Astronomy, Cardiff University, Queens Buildings, Cardiff, CF24 3AA, United Kingdom}
\email{davist@cardiff.ac.uk}

\author[orcid=0000-0003-1845-0934, gname=Toby,sname=Brown]{Toby Brown}
\affiliation{National Research Council of Canada, Herzberg Astronomy and Astrophysics Research Centre, 5071 W. Saanich Rd., Victoria, BC, V9E 2E7, Canada}
\email{tobiashenrybrown@gmail.com}

\author[orcid=0000-0001-9557-5648, gname=Amelia,sname=Fraser-McKelvie]{A. Fraser-McKelvie}
\affiliation{European Southern Observatory, Karl-Schwarzschild-Straße 2, Garching, 85748, Germany}
\email{amelia.fraser-mckelvie@eso.org}

\author[orcid=0000-0003-4569-2285, gname=Andrew ,sname=Battisti]{Andrew Battisti}
\affiliation{International Centre for Radio Astronomy Research, University of Western Australia, 7 Fairway, Crawley, 6009, Western Australia, Australia}
\affiliation{Research School of Astronomy and Astrophysics, Australian National University, Cotter Road, Weston Creek, ACT, 2611, Australia}
\email{andrew.battisti@uwa.edu.au}

\author[gname=Alessandro, sname=Boselli]{Alessandro Boselli}
\affiliation{Aix-Marseille Universit\'e, CNRS, CNES, LAM, Marseille, France}
\email{alessandro.boselli@lam.fr}

\author[orcid=0000-0002-1640-5657, gname=Pavel, sname=J\'achym]{Pavel J\'achym}
\affiliation{Astronomical Institute, Czech Academy of Sciences, Bo\v cn\'i II 1401, Prague, 14100, Czech Republic}
\email{jachym@ig.cas.cz}

\author[orcid=0000-0003-2723-0810, gname=Andrei, sname=Ristea]{Andrei Ristea}
\affiliation{Centre for Astrophysics and Supercomputing, Swinburne University of Technology, John St, Hawthorn, 3122, Victoria, Australia}
\affiliation{ARC Centre of Excellence in Optical Microcombs for Breakthrough Science (COMBS), Australia}
\email{andrei.ristea16@gmail.com}

\author[orcid=0000-0002-0956-7949, gname=Kristine, sname=Spekkens]{Kristine Spekkens}
\affiliation{Department of Physics, Engineering Physics and Astronomy, Queen’s University, Kingston, ON, K7L 3N6, Canada}
\email{kristine.spekkens@queensu.ca}

\author[orcid=0000-0003-1820-2041, gname=Sabine, sname=Thater]{Sabine Thater}
\affiliation{Department of Astrophysics, University of Vienna,
T\"urkenschanzstraße 17, 1180, Vienna, Austria}
\email{sabine.thater@univie.ac.at}

\author[orcid=0000-0001-5817-0991, gname=Christine, sname=Wilson]{Christine Wilson}
\affiliation{Department of Physics \& Astronomy, McMaster University, 1280 Main Street W, Hamilton, L8S 4M1, ON, Canada}
\email{wilson@physics.mcmaster.ca}

\begin{abstract}

The rapid quenching of satellite galaxies in dense environments is often attributed to environmental processes such as ram pressure stripping. However, stripping alone cannot fully account for the removal of dense, star-forming gas in many satellites, particularly in their inner regions. Recent models and indirect observations have suggested that star formation-driven outflows may play a critical role in expelling this remaining gas, yet direct evidence for such feedback-driven quenching remains limited. Here we report the discovery of an ionized gas outflow in NGC 4064, a Virgo cluster satellite that has already lost most of its cold gas through environmental stripping. MUSE observations from the Multiphase Astrophysics to Unveil the Virgo Environment (MAUVE) survey reveal a bi-polar outflow driven by residual, centrally concentrated star formation in NGC 4064 — despite its current star formation rate being \about 0.4 dex below the star-forming main sequence due to prior interaction with the cluster environment. The outflow’s mass loading factor is \about 2, suggesting that stellar feedback could remove the remaining gas on timescales shorter than those required for depletion by star formation alone. These results demonstrate that even modest but centrally concentrated star formation can drive efficient feedback in stripped satellites, accelerating quenching in the final stages of their evolution.

\end{abstract}

\section{Introduction}
Galaxies require a steady supply of cold gas to sustain star formation \citep[e.g.,][]{Tumlinson2017, Saintonge_2022}. In cluster satellites, this supply is rapidly disrupted by starvation, which halts fresh gas accretion \citep{larson1980, Peng2015}, and ram-pressure stripping, which removes existing gas \citep[e.g.,][]{Gunn1972, Boselli2022}. These processes effectively quench star formation in low-mass satellites (\Mst $< 10^9$ \Msun), but more massive systems often retain substantial cold gas reservoirs in their central regions even after their first pericenter passage \citep[see review by][]{Dawes2021}. Star formation-driven outflows have been proposed as a mechanism for removing the remaining gas in satellite galaxies \citep[e.g.,][]{visser25}, yet direct observational evidence for this process remains limited. Here, we present NGC 4064, a satellite caught in the act of depleting its remaining cold gas through feedback-driven outflows, illustrating how internal feedback can help complete quenching after environmentally-driven stripping.

\begin{figure}[htb!]
\centering
\includegraphics[width=0.9\textwidth]{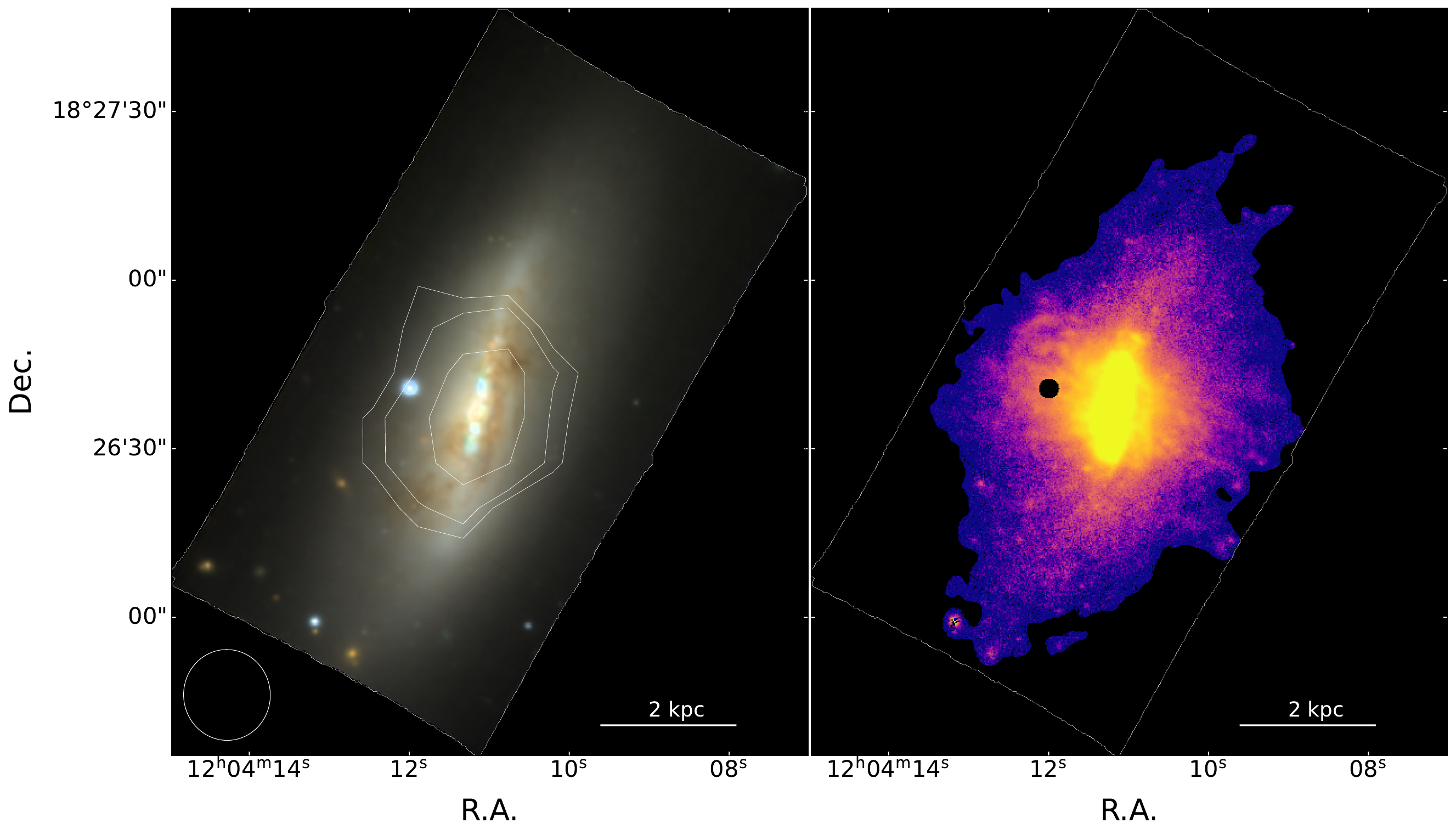}
\caption{{\bf MUSE view of NGC 4064.}
{\it Left}: Color image constructed from the $g'$, $r$, and $i$-band maps extracted from the MAUVE-MUSE cubes. White contours indicate the \hi\ distribution from \cite{Chung_2009}, with the beam size of the radio observations shown in the bottom-left corner. {\it Right}: Integrated \Ha\ flux map revealing an extended, bi-polar outflow of ionized gas emerging from the galaxy center. Note the presence of a foreground star on the eastern side that was masked. The white rectangle in both panels indicates the footprint of the MUSE mosaic.}
\label{fig1}
\end{figure}

NGC 4064 has a stellar mass of log(\Mst/\Msun) = 10.05 (from \citealt{2019_Leroy}, assuming a distance of 16.5 Mpc from \citealt{Mei_2007}) and is located in the outskirts of the Virgo Cluster, at a projected separation of 9.0\deg \citep[approximately 2.5 Mpc from M87; ][]{cortes2006}. It is classified as a `backsplash' galaxy, having previously passed within the cluster’s virial radius before exiting it \citep[]{cortes2006,2017_Yoon}. This interaction with the cluster environment has left a strong imprint on the gas content of NGC 4064: the galaxy exhibits a severely stripped atomic gas disk \citep{Chung_2009}, with its \hi\ extent reaching only 21\% of its stellar radius (measured at a surface brightness of 25 mag arcsec$^{-2}$ in $B$-band; see Fig. \ref{fig1}, left panel). This galaxy's atomic gas reservoir is roughly 60 times smaller than what is expected for an unperturbed, star-forming galaxy of the same size \citep{Chung_2009}. 

The molecular gas distribution, traced by CO(1-0), shows a bar-like morphology that is largely confined to the inner \about 15\arcsec\ \citep[\about 1.2 kpc;][]{cortes2006}. The galaxy hosts a central stellar bar, and extraplanar dust lanes extending to the south-east and north-west of the plane of the disk \citep[Fig.~\ref{fig1}, left panel;][]{cortes2006}. Over this same inner region, both the stars and molecular gas show bar-like streaming, with the CO exhibiting stronger non-circular motions and shocks, while at larger radii the stellar component becomes rotation-supported \citep{cortes2006}.

\Ha\ emission in the disk is largely confined to the central regions, with \about 90\% of the \Ha\ flux located within the inner kiloparsec, and negligible activity beyond. The star formation rate (SFR), estimated from ultraviolet (UV) and infrared (IR) emission, is 0.32 \Msunyr\ \citep{2019_Leroy}, approximately three times lower than the typical value for galaxies of comparable stellar mass on the star-forming main sequence \citep{mckelvie21}. It is estimated that the SFR was significantly reduced in this galaxy due to the interaction with the cluster environment \about 0.4 Gyrs ago \citep{crowl08}. This suppression is most likely a result of a combination of ram pressure and gravitational interactions, which has removed most of the cold gas reservoir needed to sustain star formation \citep{cortes2006}.

In this Letter, we report the discovery and characterize the properties of a bi-polar ionized gas outflow in NGC 4064 obtained with the Multi Unit Spectroscopic Explorer \citep[MUSE; ][]{Bacon2010} on the Very Large Telescope as part of the ongoing MAUVE (Multiphase Astrophysics to Unveil the Virgo Environment) survey \citep{Messenger}. We show how the outflow is likely playing a key role in the last phases of star formation quenching, suggesting that feedback can still be a primary quenching mechanism even in environmentally perturbed satellite galaxies. 
Section \ref{Section 2} describes the observations and data reduction. Section \ref{Section 3} outlines the methodology used to disentangle the outflow from the disk component and to quantify its properties. In Section \ref{Section 4}, we present the outflow geometry, kinematics, ionization state, and mass of the outflowing gas. Finally, Section \ref{Section 5} discusses the implications of our results and highlights considerations for future work.

\section{Observations and data reduction} \label{Section 2}
NGC 4064 was observed with MUSE at the Very Large Telescope as part of the MAUVE-MUSE survey (ESO Large Program 110.244E; PI: Cortese), using two pointings along the stellar disk selected to match the extent of the galaxy’s molecular gas reservoir from the Virgo Environment Traced in CO survey \citep{VERTICO}. The observations and data reduction follow the procedure detailed in \cite{ADAM}. Briefly, MUSE was operated in Wide Field Mode, providing a $\sim$ 1\arcmin $\times$ 1\arcmin\ field of view with a spatial scale of 0.2 arcsec pixel\(^{-1}\) and a spectral sampling of 1.25  \(\rm \AA\) pixel\(^{-1}\), corresponding to a velocity scale of 57 km s\(^{-1}\) at the wavelength of the \Ha\ line. The typical seeing during the observations was $\sim$0.76\arcsec. At the adopted Virgo distance of 16.5 Mpc \citep{Mei_2007}, this corresponds to a spatial resolution of $\sim$61 parsec. The science-ready cube was obtained using the \textsc{pymusepipe} reduction pipeline \citep{emsellem22} as described in \cite{ADAM}.

To extract a line-only data cube from the science-ready MUSE cube we took advantage of the new Galaxy IFU Spectroscopy Tool (nGIST) pipeline \citep{ngist,McKelvie2025}, which offers a flexible way to exploit the penalized pixel fitting \citep[\textsc{pPXF}; ][]{ppxf} code.
The cube was Voronoi binned with a target S/N=40, where the S/N ratio is computed from 4800 \AA\ to 7000 \AA. The stellar continuum was modeled by fitting the same wavelength range, while the major gas emission lines, plus sky lines and NaD absorption (both Milky Way and galaxy absorption) were masked. We used a library of 40 templates extracted from the Medium resolution INT Library of Empirical Spectra \citep[MILES;][]{miles} simple stellar population models, generated with a Chabrier \citep{chabrier_imf} initial mass function, BaSTI \citep[a Bag of Stellar Tracks and Isochrones;][]{basti} isochrones, ten ages (0.15, 0.25, 0.40, 0.70, 1.25, 2.0, 3.0, 5.0, 8.5, 14 Gyr) and four metallicities ([Z/H] = -1.5, -0.35, 0.06, 0.4). A cubic spline was then used to interpolate the best-fitting continuum model (obtained from the binned cube) across each Voronoi bin to provide data on a spaxel scale and then subtracted from the MUSE cube to obtain a line-only cube.

For the analysis presented here, we focused on the \Ha\ and [NII]\(_{6583}\) lines, creating single-line cubes and masking foreground stars for subsequent kinematic modeling. In addition, to remove any residuals due to imperfect fitting, we re-fit a second-order polynomial baseline continuum around the emission lines of interest and subtract it from the line-only cube (Fig. \ref{fig1}, right panel). \\

\noindent
\section{Methods} \label{Section 3}

\subsection{3D Kinematic Modeling of the Disk and Extraction of the Ionized Outflow}
In order to characterize the outflow in NGC 4064 and quantify its properties, we need to identify regions where the H$\alpha$ emission is dominated by outflowing gas. To achieve this, we constructed a three-dimensional (3D) model of the galaxy's large-scale rotating gas disk, noting that the inner \about20\arcsec\ (\about1.6 kpc) along the bar shows strong non-circular motions most likely dominated by the radial streaming or outflowing of gas. We then removed the modeled component from the observed cube, thereby isolating non-circular components associated with the outflow. The model was generated with the Kinematic Molecular Simulation \citep[\textsc{KinMS;}][]{KinMS} software package, which simulates gas kinematics using physically motivated analytic profiles while accounting for projection effects, beam smearing and disk thickness.The fitting was performed via the \textsc{KinMS\_fitter} interface, which uses the GAStimator package that implements a Markov Chain Monte Carlo sampler to match simulated cubes to the observed data. 

Our modeling followed a two-step approach. We first used the skySampler package \citep{Mark_skysampler}, which samples the observed \Ha\ moment-0 map to generate an input flux distribution for \textsc{KinMS}. This allows the model to reproduce the integrated intensity map by construction while modeling the disk velocity field. We assume pure rotation and parameterize the rotation curve with an arctangent function \citep[e.g.,][]{Courteau1997}, so that gas not captured by this rotation-dominated model can be attributed to non-circular motions.
While skySampler provides the best constraint to the gas kinematics, it assumes that the whole line emission belongs to the disk. Thus, we performed a second KinMS run in which the kinematic parameters of the arctangent profile were fixed to the best-fitting values from the first iteration (see Fig. \ref{fig2}, right panel). In this second step, the surface-brightness distribution was modeled analytically rather than drawn from the data, enabling a clean recovery of the disk component and isolation of the outflowing component.

We tested a range of surface brightness models: single and double exponential disks, Gaussian rings, and combinations thereof, and selected the best-fitting model using the Bayesian Information Criterion, following the procedure outlined in \cite{Hogarth_2023}. We also verified convergence in the posterior distributions of the parameters for each model. The preferred surface brightness profile is a double exponential disk that includes a warp to account for the change in position angle between the bar and inner disk. We used initial guesses for the inclination and position angles \citep{VERTICO}, which were free to vary within realistic ranges. We attempted to include an additional velocity field component to account for gas motion along the bar, but this prevented the fit parameters from converging. Since the bar region is too complex to model reliably and is excluded from our extracted outflow, this does not affect the results. The seeing from the MUSE observations was used as the effective beam size when accounting for beam smearing in the models.

\begin{figure}[h]
\centering
\includegraphics[width=\textwidth]{stellar_gas_model_mom1_FINAL.png}
\caption{{\bf The stellar, ionized gas and modeled velocity maps of NGC 4064.}
{\it Left}: stellar velocity field from the MUSE data.
{\it Centre}: \Ha\ flux-weighted velocity map from the MUSE data.
{\it Right}: Corresponding map from the best-fitting model cube of the gas disk component generated with \textsc{KinMS}.}
\label{fig2}
\end{figure}

\begin{figure}[h]
\centering
\includegraphics[width=0.9\textwidth]{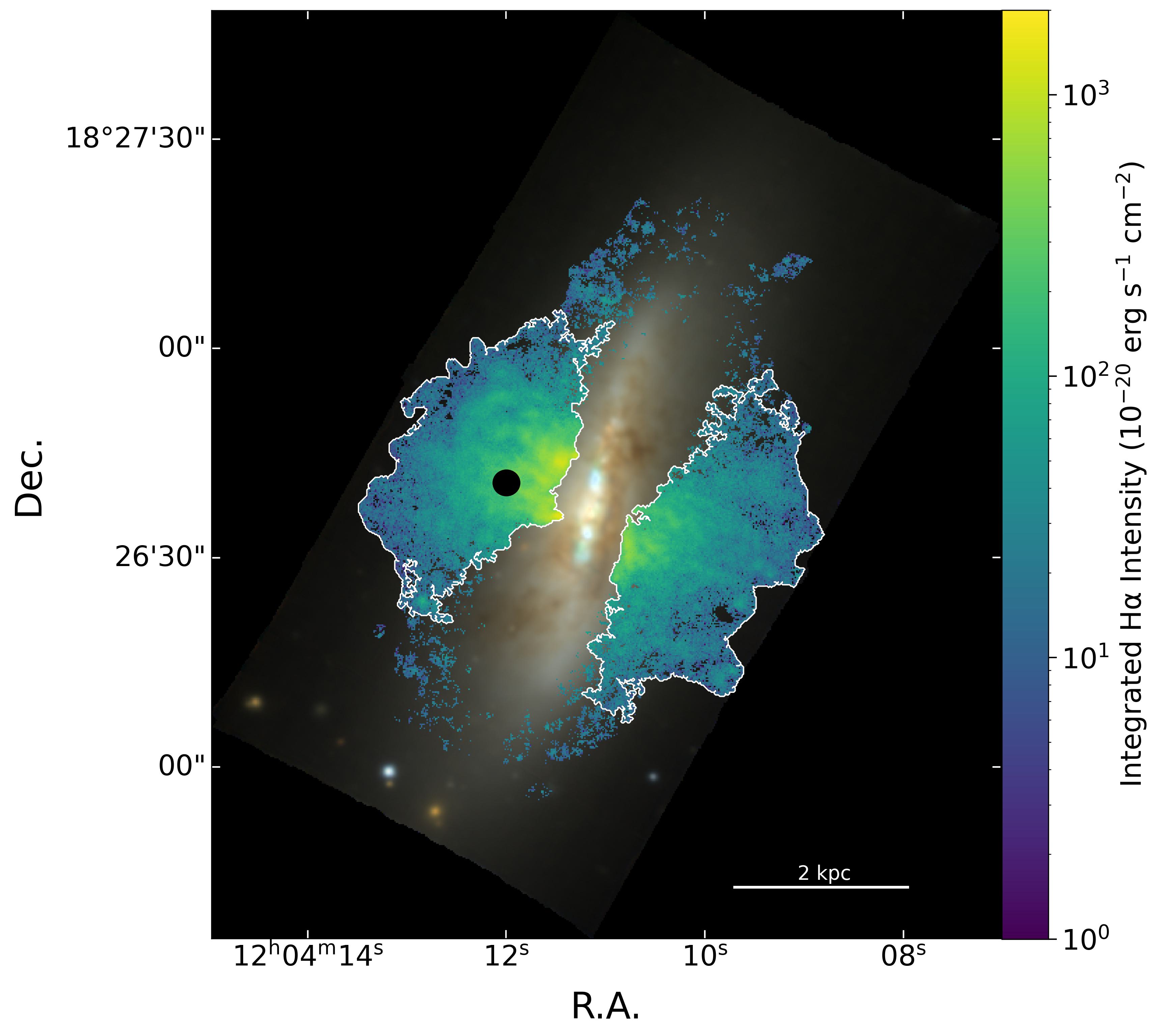}
\caption{{\bf The ionized gas outflow of NGC 4064.} 
The outflowing ionized gas, isolated by subtracting a kinematic model of the disk (Fig \ref{fig2}, right panel), is overlaid on the MUSE three-color image (same as Fig. \ref{fig1}, left panel). The white contours show the two largest contiguous regions of residual \Ha\ flux. A foreground star on the eastern side was masked.}
\label{fig3}
\end{figure}

To isolate the ionized outflow, we subtracted the best-fitting \textsc{KinMS} model cube of the gas disk from the observed \Ha\ data cube, producing a residual cube that highlights non-circular motions. We then applied a mask based on the ratio of observed-to-model flux, retaining only regions where this ratio exceeds 1.5. This threshold maximizes sensitivity to the outflow while reducing contamination from the rotating gas disk or bar-driven non-circular motions. The integrated \Ha\ intensity map of the outflow-dominated regions thus obtained is overlaid on the color image extracted from the MUSE datacube in Figure~\ref{fig3}. Due to the strong influence of the stellar bar in the central regions of NGC 4064, we are unable to disentangle outflow components from bar-driven or rotational motions in the galaxy center with high confidence. However, in the outer regions where the outflow dominates, the disk model provides a reliable representation of the kinematics of the rotating gas, as can be seen by comparing this to the stellar velocity field obtained from our MUSE data (Fig. \ref{fig2}, left), which is unaffected by the outflow. This means that we might be missing ejected gas from the inner \about 1 kpc region of the disk, thus underestimating the outflow mass -- however, this would only reinforce our main conclusions. Our approach is deliberately conservative, restricting the analysis to regions unambiguously dominated by outflowing H$\alpha$-emitting gas. For subsequent analysis, we selected the two largest contiguous regions in the residual cube -- corresponding to the eastern and western lobes of the outflow -- and defined contours enclosing 100\% of the residual \Ha\ flux within each. These spatial masks were used to emphasize the most extended and coherent structures while minimizing contributions from fragmented or ambiguous features near the disk plane (see Fig.~\ref{fig3}). \\

\section{Results} \label{Section 4}

\begin{figure}[h]
\centering
\includegraphics[width=1\textwidth]{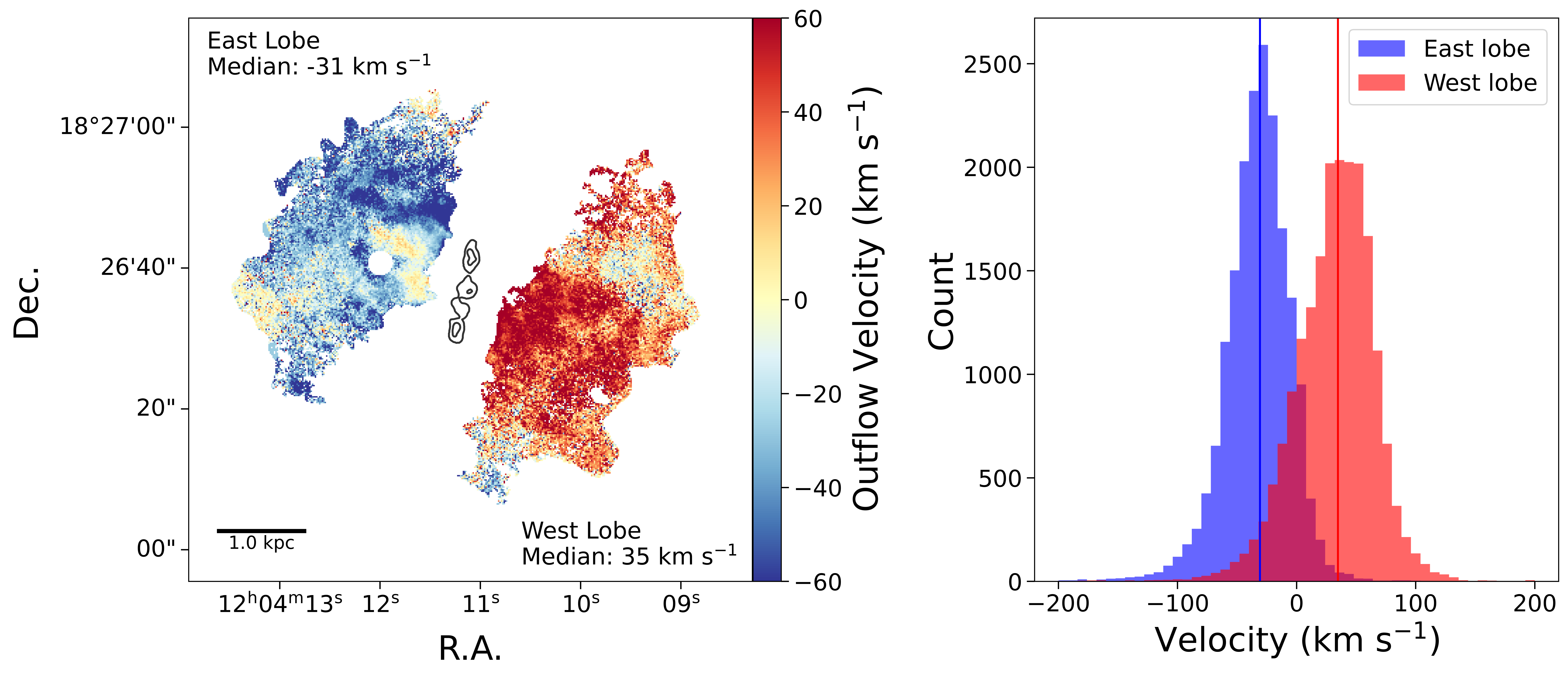}
\caption{{\bf Kinematics of the ionized outflow in NGC 4064.} {\it Left}: Line-of-sight velocity map of the \Ha\ outflow, with median velocities indicated for each lobe. Black contours trace the central stellar region as seen in the \Ha\ flux.
{\it Right}: Line-of-sight velocity distributions for the western (red) and eastern (blue) outflow lobes. Solid lines mark the median velocities.}
\label{fig4}
\end{figure}

\noindent
\subsection{Geometry, Velocity and Ionization of the Outflow}

Our MUSE observations of NGC 4064 reveal the presence of a bi-polar outflow of ionized gas emerging from the galaxy’s central region, coincident with the stellar bar (Fig.~\ref{fig3}). Previous data hinted at the presence of short filaments of ionized gas extending perpendicular to the bar \citep{cortes2006}, but the new MAUVE-MUSE observations reveal that these features are merely the brightest peaks of a larger ionized outflow extending \about 2 kpc above and below the disk of NGC 4064. The ionized gas shows a clumpy structure and reaches the edge of the MUSE pointings (at least on the western side, see Fig. \ref{fig1}), which were designed to encompass the full extent of the galaxy’s molecular gas reservoir, suggesting that the outflow may extend beyond 2 kpc.

We estimate the geometry and velocity of the outflow under the assumption that it is launched perpendicular to the plane of the disk. The projected extent of the outflow is measured as the distance above the central disk to the most distant spaxel with \Ha\ emission, for both the eastern and western lobes. These projected distances are deprojected using the galaxy’s inclination \citep[$i$ = 70\deg;][]{cortes2006}, applying a factor of 1/sin($i$). To estimate the opening angle of the outflow, we use contours derived from the \Ha\ moment-0 map to trace the boundaries of the extended emission. Assuming symmetry between the eastern and western components, straight lines are fit to the edges of the outflow in a rotated coordinate frame aligned with the galaxy’s stellar position angle (PA = 350\deg). The opening half-angle is then calculated as $\theta = 90$\deg - arctan($m$/sin($i$)), where $m$ is the slope of the fitted line. We measure an opening angle for NGC 4064's outflow of 57.7\deg. The width of the outflow base is estimated by fitting a straight line to the contours at the base of the emission and calculating the intersection with the opening-angle fits. The outflow base width, averaged over both lobes, is 0.75 kpc.

Line-of-sight velocities are measured from the flux-weighted velocity map of the outflow (Fig.~\ref{fig4}). For each lobe, we determined the median projected velocity and applied a 1/cos(i) correction to obtain the intrinsic value. This simple approximation assumes that the outflow is launched perpendicular to the disk and as such it does not capture deviations expected for gas expanding radially away from the galaxy centre. It does, however, provide a reasonable first-order estimate of the de-projected outflow motions based on the global inclination of the system.

Figure~\ref{fig4} shows the projected velocity field along with histograms of the velocity distribution in the eastern and western lobes, highlighting kinematics that are clearly distinct from the rotating disk (Fig.~\ref{fig2}). The median velocities and overall distributions are comparable on both sides, although the spatial substructure differs in detail.

To assess whether the outflowing gas can escape the galaxy’s gravitational potential, we adopt a simplified estimate for the escape velocity based on a spherical isothermal potential, \(v_{esc} \approx 3v_{cir}\)\footnote[1]{While simple, this estimate results in an escape velocity that is not strongly dependent on location in the galaxy. As noted in section 4.6. and Figure 6 of \cite{winds2005}, \(v_{esc} \approx (2.6-3.3) \times v_{cir}\) for \(r_{max}\)/r = 10-100 where \(r_{max}\) is the radius of the truncated isothermal sphere.} \citep{Veilleux2020}. Since the gas disk is truncated, we use the MUSE stellar kinematics to derive a maximum rotational velocity of \about 96 \kms\ within the MUSE field of view. Even without modeling the full rotation curve or accounting for the contribution of random motions to the galaxy’s dynamical support, this rotational velocity implies an escape velocity of \about 288 \kms, which is significantly higher than the median deprojected outflow velocity of 95.7 \kms. This suggests that the bulk of the outflowing material is unlikely to escape the galaxy’s potential well and may instead fall back or remain in the circumgalactic medium. However, while NGC 4064 currently lies beyond the Virgo cluster’s virial radius, where ram pressure is expected to be weak, its residual effects cannot be excluded, and stripping of the outflowing gas could become significant if the galaxy later falls back into the cluster.

The spatial coincidence between the extrapolated launching base of the outflow and the central star-forming region strongly suggests that stellar feedback is the primary launching mechanism. This interpretation is supported by the [NII]/\Ha\ emission-line ratio measured in the outflowing gas ($-0.4$ \(\pm\) 0.3 dex), indicating a comparatively soft ionising field from young OB stars, rather than the harder ionization levels associated to active galactic nucleus (AGN) activity. The absence of AGN signatures in the central regions of NGC 4064 from both the [NII] and [SII] Baldwin, Philips, \& Terlevich \citep[BPT;][]{BPT} diagrams derived from our MUSE spectroscopy further reinforces a star formation-driven origin. Taken together, the geometry, kinematics, and ionization state of the gas indicate that moderate but centrally concentrated star formation -- with an average SFR surface density within the inner kiloparsec of \about 0.5 \Msun~yr$^{-1}$~kpc$^{-2}$ -- is capable of driving a large-scale, coherent outflow in this satellite galaxy.

\noindent
\subsection{Ionized Gas Mass and Mass Loading Factor}
We estimate the ionized gas mass and outflow rate in NGC 4064 using the final \Ha\ outflow cube. We follow the approach described in \cite{ADAM}, based on MUSE observations of NGC~4383 taken as part of our MAUVE program. Briefly, the ionized gas mass is estimated by combining the volume of the emitting region with the electron density derived from the observed \Ha\ surface brightness, under the standard case B recombination assumption at an electron temperature T=$10^4$ K \citep{Osterbrock_1989}. We adopt a volume filling factor of $\delta = 0.01$, consistent with the analysis of NGC 4383 \citep{ADAM}, but acknowledge that this value is a large source of uncertainty in the outflow mass estimate. To estimate the volume of the outflow, we follow the method in \cite{ADAM} and multiply the area of outflow spaxels with an effective path length. We assume an effective path length through the outflow based on the geometry of a cylinder with the same volume as the biconical frustum.

The mass outflow rate is estimated by dividing the total ionized gas mass, $M_{out}$, by the dynamical timescale, calculated from the deprojected extent ($h$) and velocity ($v_{out}$) of the outflow. 
The velocity is assumed constant and set to the average of the median values for the two lobes shown in Fig. \ref{fig4}, deprojected by the galaxy inclination:

\begin{equation}
  \dot{M}_{out}[\text{M}_\odot \, \text{yr}^{-1}] = 0.53 \left(\frac{M_{out}}{1.10 \times 10^7 \, \text{M}_\odot} \right) \left( \frac{h}{2.02 \, \text{kpc}}\right)^{-1} \left( \frac{v_{\text{out}}}{95.7 \, \text{km} \, \text{s}^{-1}} \right)
\label{eq:outflowrate}
\end{equation}

We estimate an ionized gas mass outflow rate of 0.53 \Msunyr, nearly twice the galaxy's integrated UV + IR SFR (0.32 \Msunyr; the MUSE \Ha\ flux cannot be reliably used to estimate the SFR, as it is contaminated by emission from the ionized outflow). The mass loading factor, defined as $\eta =\dot{M}_{out}/\text{SFR}$, is then $\eta=1.66$. A mass loading factor greater than 1 indicates that stellar feedback removes the remaining cold gas more rapidly than star formation consumes it in the inner disk. While subject to large uncertainties due to assumptions about geometry, density, and filling factor, this value is consistent with recent estimates of ionized gas mass loading in local galaxies.

The total cold gas mass of NGC 4064, including atomic and molecular components, and the 36\% contribution of helium, is log(M\(_{gas}\)/\Msun) = 8.49 (Table~\ref{tab1}). From this, we derive a gas depletion time of 0.97 Gyr due to star formation alone, compared to 0.58 Gyr if the gas is removed by the outflow. Since ionized gas typically represents only a small fraction of the total outflowing mass in star-forming galaxies \citep[e.g.,][]{avery22}, our calculations likely underestimate the full impact of the outflow on the gas reservoir, and thus overestimate the true depletion timescale due to stellar feedback. Although the outflowing gas may not escape the galaxy’s potential well, its displacement from the disk and potential heating likely render it unavailable for further star formation, at least on the timescales required for the outflow to deplete the remaining gas reservoir.

\begin{table}[h]
\centering
\caption{Physical Properties of NGC~4064 and its Ionized Gas Outflow.}\label{tab1}%
\begin{tabular}{ll}
\toprule
Stellar Mass, log(\Mst/\Msun)        &  10.05$^1$  \\
Star Formation Rate, SFR (\Msunyr)   &  0.32$^1$ \\
Total \hi\ Mass, log(\Mhi/\Msun)     &  7.63$^2$       \\
Total \htwo\ Mass, log(\Mhtwo/\Msun) &  8.27$^3$      \\
Inclination (\deg)                   &  70$^4$         \\
\bf{Outflow Properties}:             &   \\
Total H\(\alpha\) Flux (erg s$^{-1}$ cm$^{-2}$) & 1.62 $\times$ 10$^{-12}$ \\
Half Opening Angle, $\theta$ (\deg)             & 57.7 $\pm$ 0.8$^5$  \\
         \\
Volume (kpc$^3$)                                & 49.1            \\
Total Mass, log(M$_{out}$/\Msun)                & 7.04            \\    
Outflow Rate, $\rm \dot{M}_{out}$ (\Msunyr)     & 0.53           \\   
Mass Loading Factor, $\eta$                     & 1.66             \\
\botrule
\end{tabular}
\begin{description}
    \item[References] $^1$ \cite{2019_Leroy}; $^2$ \cite{Chung_2009}; $^3$ \cite{VERTICO}, note the 36\% helium contribution is not included in this value; $^4$ \cite{cortes2006}; $^5$ Uncertainties on the derived geometry reflect a $\pm$ 5\deg\ variation in the assumed inclination.
\end{description}
\end{table}

\section{Discussion and Conclusion}  \label{Section 5}
The results presented above suggest that, in a galaxy that has already lost most of its original gas reservoir, quenching may be driven not only by the gradual consumption of gas through ongoing star formation but also, and perhaps more significantly, by feedback from residual, centrally concentrated star formation. Although stellar feedback has previously been invoked to explain full quenching in satellite galaxies \citep[e.g.,][]{Dawes2021,visser25}, direct evidence remains limited \citep[e.g.,][]{bos4254}. NGC 4064 represents the first known example of a backsplash galaxy where the ionized gas outflow rate exceeds the depletion rate from star formation, and may be even higher when accounting for the total gas mass affected by feedback.

From the star formation and outflow rates, we derive a mass loading factor of 1.66 for NGC 4064. This is comparable to values reported for local disk galaxies of the same stellar mass but significantly higher SFR. For example, \cite{Chu2025} found total ionized gas mass loading factors between 0.32 and 2.49 (0.96 on average) for a sample of 10 local starbursting galaxies with stellar masses \Mst $\geq 10^{10}$ \Msun. A study of 12 lower-mass (\Mst \about \(10^7-10^{9.3}\) \Msun) galaxies with ongoing or recent starbursts found similar (\Ha-derived) loading factors ranging from 0.2 to 7, and noted that the concentration of star formation in the central regions was a strong predictor of whether a galaxy would develop a wind \citep{McQuinn2019}. 
Using MUSE observations of 19 nearby starburst galaxies (\Mst \about \(10^7-10^{10}\) \Msun), \cite{Marasco2023} reported significantly lower loading factors, but focused specifically on the fraction of ionized gas that exceeds the escape velocity and is thus likely to leave the galaxy permanently. This distinction highlights that high mass loading does not necessarily imply efficient gas removal from the halo.

In a broader context, NGC 4064 stands out as a gas-poor, environmentally processed system already offset below the star forming main sequence, where residual, centrally concentrated star formation is still capable of driving a substantial outflow. This also implies that the local SFR surface density and not the global SFR is key for driving outflows.

In such systems undergoing quenching, feedback efficiency (measured as outflow rate per unit SFR) may be enhanced, even if much of the outflowing gas remains gravitationally bound.
In NGC 4064, the earlier stripping of its halo and outer disk gas beyond the central 2 kpc means that the outflow encounters little ambient material, allowing it to expand more freely and potentially enhancing its quenching effect on the remaining cold gas. Furthermore, tidal and ram-pressure forces during pericenter passage are expected to have removed \about30\%-50\% of the galaxy’s dark matter \citep[depending on the exact orbital stage;][]{Niemiec2019}, reducing gravitational confinement and allowing feedback to more effectively drive gas out of the disk \citep[e.g.,][]{bos4254}.

NGC 4064’s cluster history and morphology distinguish it from the more common Virgo satellites that exhibit truncated, approximately axisymmetric gas disks and comparatively modest central SFRs (e.g.,  \citealp{Koopmann2004}), indicating that outflows as strong as those observed in this galaxy may be rare among Virgo satellites. Nevertheless, the number of Virgo spirals showing evidence for outflows (either star-formation- or AGN-driven) is growing (e.g., \citealp{kenney02,chyzy06,yoshida02,bos4254}), suggesting that such processes may represent a plausible, even when not dominant, mechanism that can facilitate the final quenching of environmentally processed satellites \citep[e.g.,][]{Dawes2021,visser25}.

As the MAUVE survey continues to expand its sample across a range of environments and evolutionary stages within Virgo, it will provide a critical statistical foundation to test how such feedback-driven quenching processes operate across the cluster population. Complementary observations from Atacama Large Millimeter/submillimeter Array (ALMA) from the MAUVE-ALMA program will further characterize the cold molecular gas content and kinematics in these galaxies, offering a more complete view of how different gas phases participate in the quenching process.

\begin{acknowledgments}
We sincerely thank the anonymous reviewer for providing valuable comments and suggestions that improved this paper. This work is carried out as part of the MAUVE collaboration (https://mauve.icrar.org/) and is based on observations collected at the ESO under ESO programme 110.244E. This paper makes use of services that have been provided by AAO Data Central  (datacentral.org.au) and used the Canadian Advanced Network For Astronomy Research (CANFAR) operated in partnership by the Canadian Astronomy Data Centre and The Digital Research Alliance of Canada with support from the National Research Council of Canada the Canadian Space Agency, CANARIE and the Canadian Foundation for Innovation. L.C. acknowledges support from the Australian Research Council Discovery Project funding scheme (DP210100337).
We thank Aeree Chung, Weiguang Cui, Elisabete Da Cunha, Rongjun Huang, Yifei Jin, Zefeng Li, and Vicente Villanueva for their comments on the manuscript.
\end{acknowledgments}

\facility{VLT-MUSE}.
\software{nGIST \citep{ngist,McKelvie2025}, KinMS \citep{KinMS}, Astropy \citep{astropy:2013,astropy:2018,astropy:2022}, Numpy \citep{harris2020array}, Matplotlib \citep{Hunter:2007}}.

\bibliography{bibliography}{}
\bibliographystyle{aasjournalv7}

\end{document}